\documentclass[conference]{IEEEtran}
\IEEEoverridecommandlockouts
\usepackage{cite}
\usepackage{amsmath,amssymb,amsfonts}
\usepackage{algorithmic}
\usepackage{algorithm}  
\usepackage{graphicx}
\usepackage{textcomp}
\usepackage{xcolor}
\usepackage{multirow}
\usepackage{lipsum}
\usepackage{tikz}

\def\BibTeX{{\rm B\kern-.05em{\sc i\kern-.025em b}\kern-.08em
    T\kern-.1667em\lower.7ex\hbox{E}\kern-.125emX}}
\newcommand\copyrighttext{%
  \footnotesize \textcopyright 2020 IEEE. Personal use of this material is permitted.
  Permission from IEEE must be obtained for all other uses, in any current or future
  media, including reprinting/republishing this material for advertising or promotional
  purposes, creating new collective works, for resale or redistribution to servers or
  lists, or reuse of any copyrighted component of this work in other works.}
\newcommand\copyrightnotice{%
\begin{tikzpicture}[remember picture,overlay]
\node[anchor=south,yshift=10pt] at (current page.south) {\fbox{\parbox{\dimexpr\textwidth-\fboxsep-\fboxrule\relax}{\copyrighttext}}};
\end{tikzpicture}%
}  
\begin{document}

\title{Continuous-time finite-horizon ADP for automated vehicle controller design with high efficiency\\
{\footnotesize }
\thanks{This study is supported by International Science \& Technology Cooperation Program of China under 2019YFE0102200, Beijing NSF with JQ18010. Special thanks should be given to TOYOTA for their support on this study. Ziyu Lin and Jingliang Duan contribute equally to this work. All correspondences should be sent to S. Li with email: lisb04@gmail.com.}
}

\author{\IEEEauthorblockN{Ziyu Lin}
\IEEEauthorblockA{\textit{School of Vehicle and Mobility} \\
\textit{Tsinghua University}\\
Beijing, China \\
linzy17@mails.tsinghua.edu.cn}
\and
\IEEEauthorblockN{Jingliang Duan}
\IEEEauthorblockA{\textit{School of Vehicle and Mobility} \\
\textit{Tsinghua University}\\
Beijing, China \\
djl15@mails.tsinghua.edu.cn}
\and
\IEEEauthorblockN{Shengbo Eben Li}
\IEEEauthorblockA{\textit{School of Vehicle and Mobility} \\
\textit{Tsinghua University}\\
Beijing, China \\
lishbo@tsinghua.edu.cn}
\and
\IEEEauthorblockN{Haitong Ma}
\IEEEauthorblockA{\textit{School of Vehicle and Mobility} \\
\textit{Tsinghua University}\\
Beijing, China \\
maht19@mails.tsinghua.edu.cn }
\and
\IEEEauthorblockN{Yuming Yin}
\IEEEauthorblockA{\textit{School of Vehicle and Mobility} \\
\textit{Tsinghua University}\\
Beijing, China \\
yinyuming89@gmail.com}
\and
\IEEEauthorblockN{Bo Cheng}
\IEEEauthorblockA{\textit{School of Vehicle and Mobility} \\
\textit{Tsinghua University}\\
Beijing, China \\
chengbo@tsinghua.edu.cn}}

\maketitle
\copyrightnotice

\begin{abstract}
The design of an automated vehicle controller can be generally formulated into an optimal control problem. This paper proposes a continuous-time finite-horizon approximate dynamic programming (ADP) method, which can synthesis off-line near-optimal control policy with analytical vehicle dynamics. Lying on the general Policy Iteration framework, it employs value and policy neural networks to approximate the mappings from the system states to value function and control inputs, respectively. The proposed method can converge to the near-optimal solution of the finite-horizon Hamilton-Jacobi-Bellman (HJB) equation. We further applied our algorithm to the simulation of automated vehicle control for the path tracking maneuver. The results suggest that the proposed ADP method can obtain the near-optimal policy with 1$\%$ error and less calculation time. What is more, the proposed ADP algorithm is also suitable for nonlinear control systems, where ADP is almost 500 times faster than the nonlinear MPC ipopt solver.
\end{abstract}


\section{Introduction}
Automated vehicles have promising potentials in safety, fuel economy, and traffic efficiency \cite{burns2013vision,duan2020hierarchical}. As one of the critical components in the automated driving systems, motion control directly dominates a vehicle's driving performance. At present, the optimal motion controller has become an indispensable capability for high-level automated vehicles. Considering the capability in handling system nonlinearities and constraints, Model Predictive Control (MPC) is the top of the list for controller design \cite{li2010model}. MPC predicts the future evolution of the system and solves the control inputs online in a receding horizon fashion.

Consequently, MPC has received growing attention in the field of autonomous driving. Borrelli \emph{et al}. (2005) proposed the nonlinear MPC control framework to perform the double lane change with active steering based on NPSOL software package. The simulation shows that under the condition of real-time computation, MPC can only obtain suboptimal solutions\cite{borrelli2005mpc}. Verschueren \emph{et al}. (2014) built on the generalized Gauss-Newton method to handle the real-time nonlinear MPC and got calculation time by an experiment on a Real-time Debian system \cite{verschueren2014towards}. Borrelli \emph{et al}. (2014) presented the robust invariant set to satisfy robust MPC constraints of states and inputs, and did experiment on dSPACE \cite{gao2014robust}. Gerdes \emph{et al}. (2016) proposed MPC with the constraint-based approach of envelope control, which ensures vehicle safety and brings the shared control scheme \cite{erlien2015shared}. Siampis (2017) solved the nonlinear MPC problem with the Real-Time Iteration scheme and the Primal-Dual Interior-Point method, and experimented on dSPACE too \cite{siampis2017real}. Kayacan (2018) presented a real-time fast iteration scheme implemented for the nonlinear MPC framework and compared the performance and computational complexity with a linear MPC framework \cite{kayacan2018experimental}. Batkovic \emph{et al}. (2019) took the constraints of moving pedestrians into consideration for motion planning and control with MPC, which was also validated on Volvo XC90 \cite{batkovic2019real}. This algorithm sacrifices solving precision to ensure real-time performance.

Existing MPC optimizers usually adopt the online calculation, which can not satisfy policy accuracy and the millisecond-level time requirements of onboard standard controllers, simultaneously. Taking the computing power of the Audi A8 L3 level autopilot controller, zFAS, as the benchmark, the calculation time of MPC algorithms mentioned above (\cite{gao2014robust,batkovic2019real},\cite{siampis2017real},\cite{kayacan2018experimental},\cite{verschueren2014towards} ) is 203.4 $\mathrm{ms}$, 103.6 $\mathrm{ms}$, 98.9 $\mathrm{ms}$, 98.4 $\mathrm{ms}$, and 41.6 $\mathrm{ms}$ respectively. In general, the computing time assigned to the automated vehicle motion control task is less than 10ms due to the limited computing capability. The fastest single-step computing time of the five MPC algorithms listed is 41.6 $\mathrm{ms}$, which is obviously difficult to meet the controller's real-time requirements. 

At present, the methods with off-line calculation have been expected to improve MPC calculation time. Approximate Dynamic Programming (ADP) method seeks approximated control policy off-line, and then apply it for online execution as a low-dimensional parameterized function \cite{werbos1992approximate}. The usage of ADP could significantly reduce the online computing burden. Its basic principle is to solve Hamilton-Jacobi-Bellman (HJB) equation off-line by implementing the policy iteration framework \cite{werbos1987building,duan2019generalized,duan2019deep}. Meanwhile, existing methods transform continuous-time systems into discrete-time families, represented by MPC. Practically, the vehicle system works in a continuous-time way, which preserves significant merits in theoretical analysis and controller synthesis.  In conclusion, solving the continuous-time ﬁnite horizon optimal control problem analytically with high efficiency becomes an urgent task.  However, there are few existing ADP techniques on solving the optimal solution for finite-horizon control, especially for continuous-time systems. It is intractable to solve the nonlinear partial differential HJB equation.

This paper proposes a continuous-time finite-horizon ADP algorithm as the fast computation technique for automated vehicles’ motion control. It can handle the general nonlinear systems with known dynamic. It employs value and policy neural networks to approximate the mappings from the system states to value function and control inputs, respectively, lying on the policy iteration framework.  We demonstrate our algorithm with simulations on the vehicle-tracking control problem.  The main contribution can be summarized as follows:

(1) We propose a finite-horizon variant of ADP algorithm, which can be used to find a nearly optimal policy off-line for a given continuous-time nonlinear system.

(2) The online operation efficiency of the off-line policy is 500 times faster than the nonlinear MPC solver.

The paper is organized as follows. Section II provides the formulation of the general optimal control problem for the vehicle controller, followed by the derivate of the reshaped HJB equation. Section III introduces the general finite-time continuous-time ADP framework with parameterized function. Section IV demonstrates the effectiveness by the simulations with a linear-quadratic system and a nonlinear system. Section V concludes the paper.
\section{PROBLEM FORMULATION}
In this section, we introduce the linear vehicle model used in the vehicle controller design, and then the formulation for the vehicle-tracking controller is built. Subsequently, we propose the finite-horizon version of the HJB equation.
\subsection{Vehicle model}
\begin{figure}[htbp]
\centerline{\includegraphics[width=0.45\linewidth]{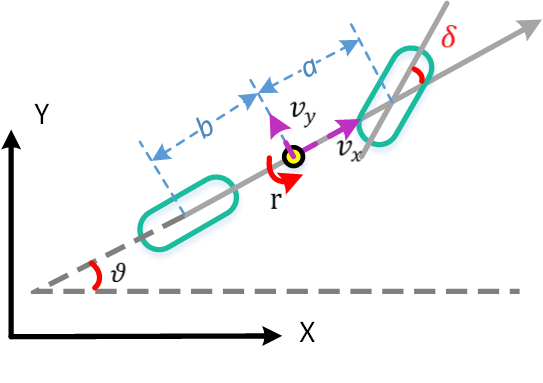}}
\caption{The simplified vehicle dynamical model.}
\label{fig1}
\end{figure}

The two degrees-of-freedom linear model is adopted to describe the vehicle dynamic shown in Fig.~\ref{fig1}. The continuous-time vehicle dynamic equations in an inertial frame can be derived as: 
\begin{equation}
\begin{aligned}
\dot{\vartheta}&=r, \\
\dot{r}&=\frac{a^{2} k_{1}+b^{2} k_{2}}{I_{z z} v_{x}} r+\frac{a k_{1}-b k_{2}}{I_{z z} v_{x}} v_{y}-\frac{a k_{1}}{I_{z z}} \delta_{,} \\
\dot{v_{y}}&=\left(\frac{a k_{1}-b k_{2}}{m v_{x}}-v_{x}\right) r+\frac{k_{1}+k_{2}}{m v_{x}} v_{y}-\frac{k_{1}}{m} \delta.
\end{aligned}
\end{equation}
Mathematical notations are listed in Table I.
\begin{table}[htbp]
\caption{Vehicle Parameters Notation}
\begin{center}
\begin{tabular}{|l|c|}
\hline Inertial heading angle & $\vartheta$ \\
\hline Yaw rate at the center of gravity (CG) & $r$ \\
\hline Constant longitudinal velocity & $v_{x}$ \\
\hline Distance from mass center to front \& rear axle & $a, b$ \\
\hline Front and rear wheel cornering stiffness & $k_{1}, k_{2}$ \\
\hline Mass & $m$ \\
\hline Moment of inertia around the z-axis & $I_{z z}$ \\
\hline
\end{tabular}
\label{tab1}
\end{center}
\end{table}

\begin{figure}[htbp]
\centerline{\includegraphics[width=0.55\linewidth]{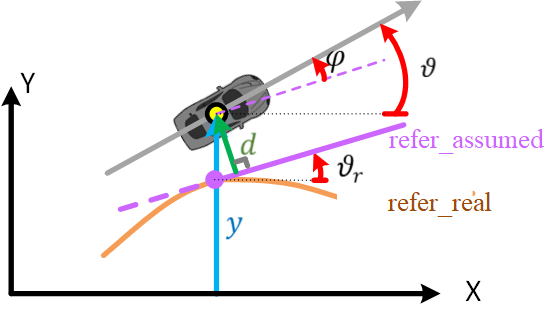}}
\caption{Illustration of position and heading angle error.}
\label{fig2}
\end{figure}

This classical bicycle model formulates the vehicle dynamics in reference to an initial coordinate system. However, for the vehicle tracking control problem, it is favorable to predict future vehicle motion considering the known reference trajectory. Hence, we introduce two new system states shown in Fig.~\ref{fig2}, which are the  position error $d$ and heading angle error $\varphi$ with respect to the assumed trajectory in the inertial coordinates:
\begin{equation}\begin{aligned}
d&=\left(y-y_{r}\right) \cos \vartheta_{r}, \\
\varphi &=\vartheta-\vartheta_{r},
\end{aligned}\end{equation}
where position error $d$ is the distance between the center of gravity (CG) and the assumed reference trajectory. It should be noted that the assumed reference trajectory is the tangent line passing the current vehicle preview point $\left(x, y_{r}(x)\right)$. $y_{r}$ and $\theta_{r}$ are $Y$ position and heading angle of the current preview point on reference trajectory, respectively. $y_{r}$ is a reference trajectory function depending on vehicle position $x$. $x, y$ are the coordinates of the vehicle $\mathrm{CG}$ in an absolute inertial frame $(\mathrm{X}, \mathrm{Y}) .$ Therefore, the vehicle absolute position $y$ is given by:
\begin{equation}y=y_{r}+d / \cos \vartheta_{r}\end{equation}

With $\dot{\vartheta}_{r}=0,$ the following differential equations are satisfied:
\begin{equation}\begin{aligned}
\dot{d}&=v_{x} \sin \varphi+v_{y} \cos \varphi, \\
\dot{\varphi}&=\dot{\vartheta}-\dot{\vartheta}_{r}=r.
\end{aligned}\end{equation}

We consider the vehicle trajectory tracking control problem with linear vehicle dynamic. The system states are $\left[d, \varphi, r, v_{y}\right]$ and control input is $\delta,$ listed in Table II. We assume the system is stable when $x$ belongs to compact set $\Omega$. With the assumption that $\varphi$ is small, the  bicycle model's lateral dynamic in the inertial frame is described as follows:
\begin{equation}\begin{aligned}
x&=\left[\begin{array}{c}
d \\
\varphi \\
r \\
v_{y}
\end{array}\right], \dot{x}=A x+B u, B=\left[\begin{array}{c}
0 \\
0 \\
-\frac{a k_{1}}{I_{z z}} \\
-\frac{k_{1}}{m}
\end{array}\right], \\
A&=\left[\begin{array}{cccc}
0 & v_{x} & 0 & 1 \\
0 & 0 & 1 & 0 \\
0 & 0 & \frac{a^{2} k_{1}+b^{2} k_{2}}{I_{z z}} & \frac{a k_{1}-b k_{2}}{I_{z z} v_{x}} \\
0 & 0 & \frac{a k_{1}-\tilde{b k_{2}}}{m v_{x}}-v_{x} & \frac{k_{1}+k_{2}}{m v_{x}}.
\end{array}\right]
\end{aligned}\end{equation}
\begin{table}[htbp]
\caption{State And Control Input}
\begin{center}
\begin{tabular}{|c|c|c|}
\hline 
\multirow{2}*{ State }& Distance between CG \& trajectory & $d$ \\
\cline { 2 - 3 } & The heading angle between the vehicle \& trajectory & $\varphi$ \\
\cline { 2 - 3 } & Yaw rate at the center of gravity (CG) & $r$ \\
\cline { 2 - 3 } & Lateral velocity & $v_{y}$ \\
\hline Input & Steer angle of front wheel & $\delta$ \\
\hline
\end{tabular}
\label{tab2}
\end{center}
\end{table}
\subsection{Vehicle tracking problem}

Generally, the design of the automated vehicle controller can be formulated into an optimal continuous-time control problem. The basic idea of MPC is to use the vehicle dynamic model to predict the vehicle tracking system's future motion. Based on the prediction, at each time step $t$, the utility function is designed to minimize the tracking errors:
\begin{equation}l(x(\tau), u(\tau))=Q d^{2}+R \delta^{2},\end{equation}
where $Q$ and $R$ denote penalty coefficient of distance error and steer angle, respectively.

Hence, the optimal control problem (OCP) for $\forall x \in \Omega$, $\forall t \in$
$[0, T]$ is given by:
\begin{equation}\begin{aligned}
V^{*}(x(t), t)&=\min _{u} V(x, t)=\min _{u} \int_{t}^{T} l(x(\tau), u(\tau)) d \tau\\
&\text { s.t. } \quad \dot{x}=f(x, u),
\end{aligned}\label{eq7}\end{equation}
where $T$ is the final time, $t \in\lceil 0, T\rceil$ is initial time, $\tau$ is the virtual time in the future horizon, $u \in \mathbb{R}^{n}$ is the control input (action, in other words), $x \in \Omega \subset\mathbb{R}^{m}$ is the state, $V$ is the state value function, $l(\cdot,\cdot)$ is the utility function $(l(\cdot,\cdot)>0$ except for the equilibrium).
\subsection{Reshape of HJB equation}

The basic principle of continuous-time finite-horizon OCP is to seek a policy $u=\pi(x,t)$ to minimize $V(x,t)$ for $\forall x$. For this purpose, we first introduce the Hamilton function:
\begin{equation}
H(x,u,V^{\pi}(x,t))=l(x, u)+\frac{\partial V^{\pi}(x, t)}{\partial x^{\top}}f(x,u).
\end{equation}

The Hamiltonian satisfies the following the finite-horizon self-consistency condition
\begin{equation}
\begin{aligned}
H(x,u,V^{\pi}(x,t))=-\frac{\partial {V^{\pi}(x, t)}}{\partial t},\quad \forall t \in[0, T].
\end{aligned}
\label{eq9}
\end{equation}

We have deduced that $\frac{\partial V^{\pi}(x, t)}{\partial t}=-l\left(x(T), u(T)\right)$, where $x(T)$ and $u(T)$ are obtained from $x(t)$ according to policy $\pi(x,t)$. The optimal solution satisfies the following finite-horizon HJB equation:
\begin{equation}
\min _{u}\left\{l(x, u)+\frac{\partial V^{*}(x, t)}{\partial x^{\top}} f(x, u)\right\}=l\left(x^{*}(T), u^{*}(T)\right).
\label{eq8}
\end{equation}

The optimal control $\pi^{*}(x, t)$ for $\forall x \in \Omega$, $\forall t \in$
$[0, T]$ can be derived as
\begin{equation}\pi^{*}(x, t)=\arg \min _{u} H\left(x, u, \frac{\partial V^{*}(x, t)}{\partial x^{\top}}\right).\label{eq10}\end{equation}

It is known that the optimal control policy for the vehicle control system is the solution of finite-horizon HJB equation \eqref{eq8}, which is the sufficient and necessary condition of optimality. There are two variables $V^{*}(x, t)$ and $\pi^*(x,t)$. However, HJB equation is a typical nonlinear partial differential equation, which is generally difficult to be analytically solved.
\section{FINITE-HORIZON ADP ALGORITHM}
In this section, we propose the continuous-time (CT) finite-horizon ADP algorithm. It is an iterative algorithm to find the nearly optimal solution of the HJB equation \eqref{eq8}, which involves computation cycles between policy evaluation based on \eqref{eq9} and policy improvement \eqref{eq10}.  The value function and policy approximation can be parameterized by neural networks.
\subsection{Basic principle}

The formulation of MPC is a useful tool to understand the basic principle of ADP. MPC and ADP share very similar receding horizon control procedure, yet differs on how to find optimal action. ADP first computes optimal policy off-line and then applies it online. Differently, MPC calculates optimal action in the way of receding horizon optimization. Hence, ADP can efficiently reduce online computation since it moves most of the computation to the off-line stage. 

\begin{figure}[htbp]
\centerline{\includegraphics[width=0.6\linewidth]{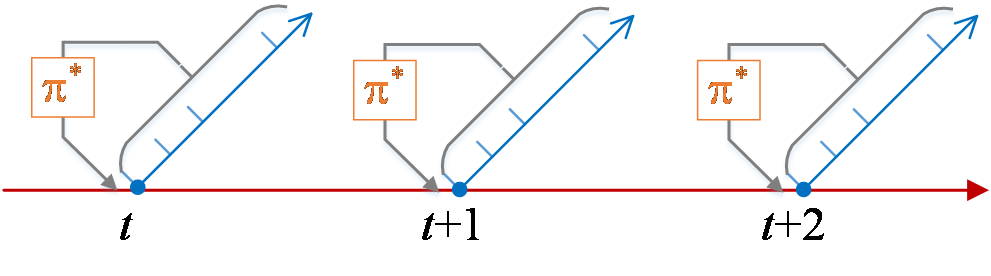}}
\caption{Receding horizon control for ADP execution.}
\label{fig3}
\end{figure}

We can always divide the implementation of OCP into two steps: (a) objective optimization, and (b) action implementation. ADP minimizes the objective function off-line and implements optimal action online. This two-step procedure can also be understood from the viewpoint of receding horizon control, i.e., define a problem in virtual time and use optimal policy in real time, shown in Fig.~\ref{fig3}. 

For this algorithm, at each time step $t$, a performance index is optimized with respect to a sequence of future steering moves in order to follow the given trajectory. The first of such optimal moves is the control action applied to the plant at time $t$. At time $t+1$, a new optimization is solved over a shifted prediction horizon.
\subsection{Iterative ADP Framework}

\begin{figure}[htbp]
\centerline{\includegraphics[width=1\linewidth]{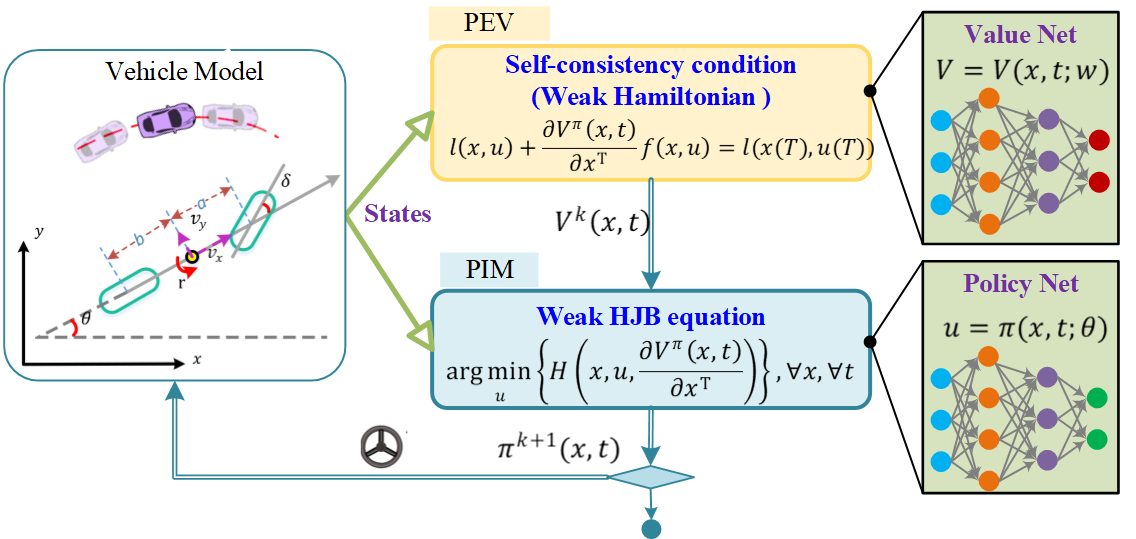}}
\caption{CT finte-horizon ADP with parameterized function}
\label{fig4}
\end{figure}

The key point for the proposed algorithm is to solve HJB equation \eqref{eq8} iteratively. Lying on general policy iteration framework, it iteratively performs two steps shown in Fig.~\ref{fig4}:

(1) Policy Evaluation (PEV)

PEV is to solve the value function $V^{\pi}(x, t)$, which is the solution of the self-consistency condition in (9). Given a policy  $\pi^{k}(\cdot)$ , the  $V^{\pi}(x, t)$ can be solved by 
\begin{equation}H\left(x, u, \frac{\partial V^{\pi}(x, t)}{\partial  x^{\top}}\right)=l(x(T), u(T)).\end{equation}

In addition, $V^{\pi}(0)=0$ is an initial condition when solving
(8). Here, Hamiltonian $H$ is a little different from true Hamiltonian, in the fact that optimal value $V^{*}(x, t)$ is replaced with estimated value $V^{\pi}(x, t)$.

(2) Policy Improvement (PIM)

PIM is to find a better policy $\pi^{k+1}(x, t)$ by minimizing the "weak" HJB equation. Here, "weak" means that optimal $V^{*}(x, t)$ in true Hamiltonian is replaced by estimated $V^{\pi}(x, t)$ at each ADP iteration.
\begin{equation}\pi^{k+1}(x, t)=\arg \min _{u}\left\{H\left(x, u, \frac{\partial V^{\pi}(x, t)}{\partial x^{\top}}\right)\right\}\end{equation}
where $V^{\pi}(\cdot)$, $x, t$ are known, and $\pi^{k+1}(\cdot)$ is unknown to be solved.
\subsection{Value Function and Policy Approximation}

To get the nearly optimal solution quickly, both value function and policy are parameterized by the approximation functions: 
\begin{equation}\begin{aligned}
V(x, t) &\cong V(x, t ; w), \\
u &\cong \pi(x, t ; \theta).
\end{aligned}\end{equation}

The value function (i.e., critic) is a parameterized function with parameter $w$, and the policy (i.e., actor) is a parameterized function with parameter $\theta$. Value function and policy are updated by designing the loss functions using the gradient descent method. The operation of each approximate function is described below.

1)	Parameterized Value Function Updating rule

The critic is to minimize the following average square error based on self-consistency condition at k-step iteration PEV:
\begin{equation}
J_{\text {Critic }}=\underset{x,t \sim d_{x,t}}{\mathbb{E}}\left\{\frac{1}{2}(H-l(x(T), u(T)))^{2}\right\}.
\end{equation}
where $d_{x,_t}$ denotes the state-time pair distribution over $x \in\Omega$ and $t \in\lceil 0, T\rceil$. The true-gradient of critic function equals to:
\begin{equation}
\frac{\partial J_{\text {Critic }}}{\partial w}=\underset{x,t \sim d_{x,t}}{\mathbb{E}}\left\{(H-l(x(T), u(T))) \frac{\partial\left(\frac{\partial V_{w_{k}}}{\partial x^{\top}} f\right)}{\partial w}\right\}.
\label{eq16}\end{equation}

2)	Parameterized Policy Updating rule

The parameterized actor is to minimize “weak” HJB equation: 
\begin{equation}
\begin{aligned}
J_{\text {Actor}}&=\underset{x,t \sim d_{x,t}}{\mathbb{E}}\left\{H\left(x, \pi(x,t ; \theta^{k}), \frac{\partial V\left(x, t ; w^{k}\right)}{\partial x^{\top}}\right)\right\} \\
&=\underset{x,t \sim d_{x,t}}{\mathbb{E}}\left\{l\left(x, \pi(x,t ; \theta^{k})\right)+\frac{\partial V_{w^{k}}}{\partial x^{\top}} f\left(x, \pi(x, t ; \theta^{k})\right)\right\}.
\end{aligned}
\end{equation}

The gradient of $J_{\text {Actor}}$ is
\begin{equation}
\frac{\partial J_{\text {Actor}}}{\partial \theta}=\underset{x,t \sim d_{x,t}}{\mathbb{E}}\left\{\left(\frac{\partial l(x, u)}{\partial u}+\frac{\partial\left(\frac{\partial V_{^{w^{k}}}}{\partial x^{\top}} f\right)}{\partial u}\right) \frac{\partial \pi(x, t ; \theta^k)}{\partial \theta}\right\}.
\label{eq18}\end{equation}

The neural networks can be updated with the given update gradients, the pseudo-code of ADP in GPI framewok is shown in Algorithm 1. The loop of PEV and PIM iteratively solve the parameterized value function $V(x, t ; w)$ and parameterized policy $\pi(x, t ; \theta) .$ The iteration of PEV and PIM will switch gradually to the HJB solution. In other words, Algorithm 1 will iteratively converge to the value function, $V\left(x, t ; w^{*}\right)=V^{*}\left(x, t\right),$ and optimal control policy $\pi\left(x, t; \theta^{*}\right)=$ $\pi^{*}\left(x, t \right)$. 

\begin{algorithm}
\caption{CT Finite-horizon ADP}
\label{alg:CT Finite-horizon ADP}
\begin{algorithmic}
\STATE {Initial with arbitrary $w^{0},\theta^{0}$, learning rates $\alpha_{w}$ and $\alpha_{\theta}$} 
\REPEAT
\STATE Rollout from $\forall x_t \in \Omega$ with policy $\pi_{\theta^{k}}$
\STATE Receive and store $x(T)$ 
\STATE Step 1 Policy evaluation: 
\STATE Calculate $\frac{\partial J_{\text{Critic}}}{\partial w}$ using \eqref{eq16}
\STATE Update value function using
\STATE \begin{equation} w^{k}=w^{k}-\alpha_{w} \frac{\partial J_{\text {Critic }}}{\partial w}\end{equation}
\STATE Step 2 Policy improvement: 
\STATE Calculate $\frac{\partial J_{\text{Actor}}}{\partial \theta}$ using \eqref{eq18}
\STATE Update policy using
\STATE \begin{equation}\theta^{k}=\theta^{k}-\alpha_{\theta} \frac{\partial J_{\text {Actor }}}{\partial \theta} \end{equation}
\UNTIL{$w,\theta$ converge}
\end{algorithmic}
\end{algorithm}

\section{SIMULATION AND RESULTS}
\begin{figure}[htbp]
\centerline{\includegraphics[width=0.45\linewidth]{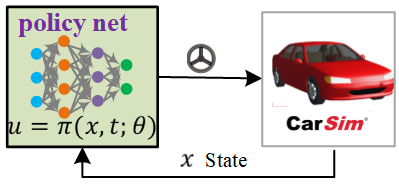}}
\caption{Framework of Carsim verification.}
\label{fig5}
\end{figure}

To support the continuous-time finite-horizon ADP algorithm, we apply it to the simulations with linear vehicle dynamic. It is used to solve the nearly optimal policy for the vehicle tracking system. CarSim supplies the vehicle model with steering input from the controller, shown in Fig.~\ref{fig5}. The simulation results show that it achieves tracking the desired trajectory. Meanwhile, for a linear problem, the algorithm works as better as the traditional MPC method. However, it can be calculated off-line, which has a promising application prospect.

\subsection{Problem Parameters and Algorithm Details}
Consider the vehicle tracking problem with linear dynamic. The desired vehicle trajectory is shown in Fig.~\ref{fig8}. The detailed vehicle parameters are listed in Table III. The vehicle is controlled by a saturating actuator, where $\delta \in\lceil-0.35,0.35]$ rad. The vehicle longitudinal velocity $v_{x}$ is chosen as 15 $\mathrm{m} / \mathrm{s}$.

\begin{table}[htbp]
\caption{Vehicle Parameters}
\begin{center}
\begin{tabular}{|c|c|c|c|c|c|}
\hline$k_{1}$ & -88000 & $\mathrm{N} / \mathrm{rad}$ & $k_{2}$ & -94000 & $\mathrm{N} / \mathrm{rad}$ \\
\hline$a$ & 1.14 & $\mathrm{m}$ & $b$ & 1.4 & $\mathrm{m}$ \\
\hline$m$ & 1500 & $\mathrm{kg}$ & $I_{z z}$ & 2420 & $\mathrm{kg} \cdot \mathrm{m}^{2}$ \\
\hline
\end{tabular}
\end{center}
\end{table}

We further set the parameters in optimal control formulation, which include: the final time $T$ is $0.5$ $\mathrm{s}$; the weighting matrices: $Q=0.4,$ and $R=280 .$ To approximate $V(x, t ; w)$ and $\pi(x, t ; \theta),$ the 3-layer fully-connected NNs with $2^{5}$ units per layer are used. For each network, the state and $t$ input layer is followed by one hidden layer. The exponential linear units (ELUs) are chosen as activation functions. The output of value networks is $V(x, t; w)$ with softplus as the activation function. The output of policy networks is $\pi(x, t; \theta)$ with tanh function multiplied by the vector $[0.35]$, which will satisfy the input steering angle constraint. The learning rate $\alpha_{w}$ and $\alpha_{\theta}$ are both set to 0.001. We use Adam method to update the network, and the Adam update rule is used to minimize the loss functions.
\subsection{Training Results on accuracy and efficiency}

For the LQ problem, we can get the analytical solution for the discrete-time system, optimal policy, in other words. 
\begin{equation}\begin{aligned}
U^{*} &=\left[u_{0 \mid t}^{*}, u_{1 \mid t}^{*}, \cdots, u_{T-1 \mid t}^{*}\right]^{2}=-H^{-1}+F x_{t} ,\\
\end{aligned}\end{equation}
where, \begin{equation}
\nonumber
\begin{aligned}
H &=2\left(\bar{R}+\bar{S}^{\top} \bar{Q} \bar{S}\right), \quad F^{\top}=2 \bar{T}^{\top} \bar{Q} \bar{S}, \\
\bar{S} &=\left[\begin{array}{cccc}
B &  0 & \cdots & 0 \\
A B & B & \cdots& 0 \\
\cdots & \cdots & \cdots & \cdots\\
A^{N-1} B & A^{N-2} B & \cdots & B
\end{array}\right], \bar{T}=\left[\begin{array}{c}
A \\
A^{2} \\
\cdots \\
A^{N}
\end{array}\right], \\
\bar{Q} &=\left[\begin{array}{cccc}
Q & 0 & \cdots & Q \\
0 & Q & \cdots & 0 \\
\cdots & \cdots & \cdots & \cdots \\
0 & 0 & \cdots & P
\end{array}\right], \bar{R}=\left[\begin{array}{cccc}
R & 0 & \cdots & Q \\
0 & R & \cdots & 0 \\
\cdots & \cdots & \cdots & \cdots \\
0 & 0 & \cdots & R
\end{array}\right].
\end{aligned}\end{equation}

The relative policy error is obtained by comparing with optimal solution $\pi^{*}(x,t)=u_{0 \mid t}^{*}$ with the equation:
\begin{equation}e_{\pi}=\underset{x,t \sim d_{x,t}}{\mathbb{E}} \left[\frac{\pi(x,t ; \theta)-\pi^{*}(x,t)}{\underset{x,t \sim d_{x,t}}{\max} \pi^{*}(x,t)-{\underset{x,t \sim d_{x,t}}{\min}} \pi^{*}(x,t)}\right].\end{equation}

\begin{figure}[htbp]
\centerline{\includegraphics[width=0.7\linewidth]{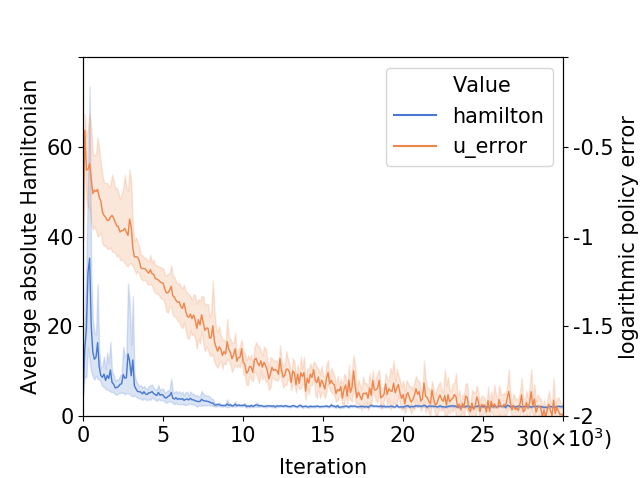}}
\caption{Policy error and Hamilton of the Algorithm}
\label{fig6}
\end{figure}

The test set contains 500 states randomly selected from the compact set $\Omega$ at the beginning of each simulation. The algorithm was run ten times, and the mean of the training performance is shown in Fig.~\ref{fig6}. We plot the policy error compared with MPC optimal solution and Hamilton of the Algorithm at each iteration. After 30000 iterations, the policy error is less than 1$\%$, shown in Fig. 6. The result indicates that the algorithm can converge policy to optimality.

\begin{figure}[htbp]
\centerline{\includegraphics[width=0.85\linewidth]{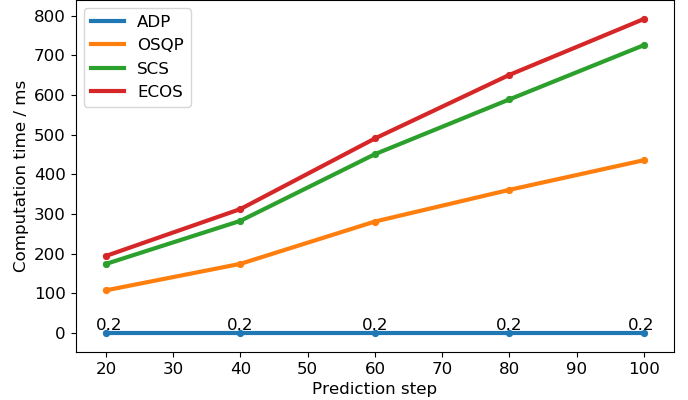}}
\caption{Comparison of computation time of ADP and QP solvers}
\label{fig7}
\end{figure}

For this optimal control problem with a linear model, we can transfer it into the linear-quadratic programming (QP) problem, and then solve it with some QP solvers, such as OSQP \cite{stellato2020osqp}, SCS \cite{o2016conic} and \cite{domahidi2013ecos}. The simulations are carried on the personal computer with i5-8250U CPU. Fig.~\ref{fig7} compares the calculation efﬁciency of ADP and the QP solvers based on the modeling language CVXPY \cite{diamond2016cvxpy} under different online applications' prediction steps. The calculation time of the QP solvers is not only longer than ADP under the same horizon but also increases faster as the number of predicted steps increases. Speciﬁcally, when $N=100$, the calculation of the proposed ADP is about 2000 times faster than the fastest QP solver OSQP. This demonstrates the effectiveness of the ADP method in linear systems control.

\subsection{Test results on tracking problem}

\begin{figure}[htbp]
\centerline{\includegraphics[width=0.75\linewidth]{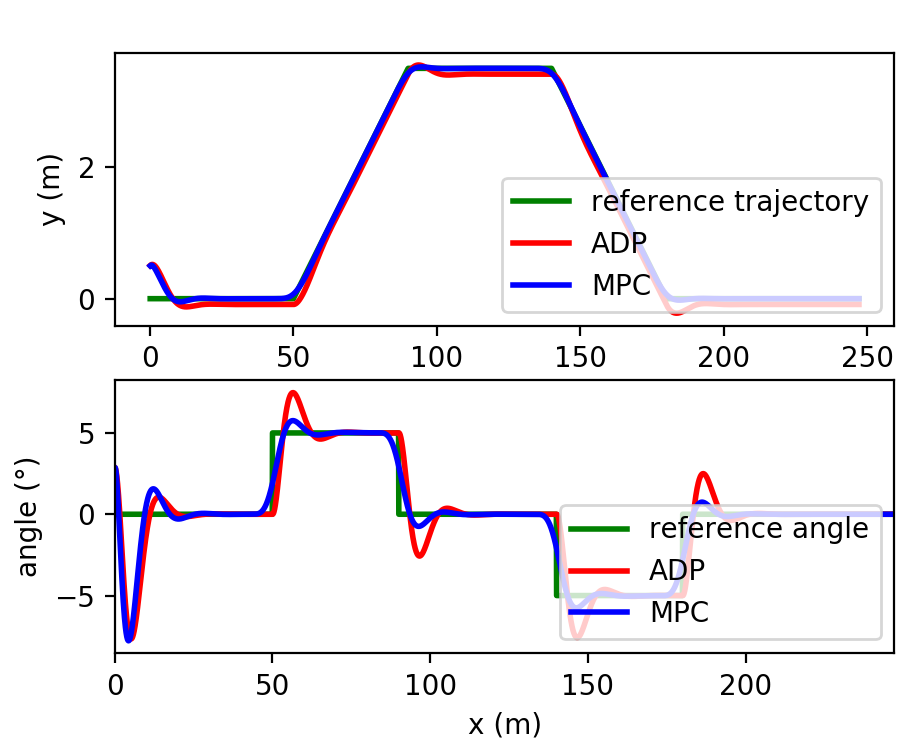}}
\caption{Trajectory comparison with linear dynamic}
\label{fig8}
\end{figure}

The vehicle trajectory controlled by the trained finite-horizon ADP algorithm is shown in Fig.~\ref{fig8}. The learned policy can achieve following the desired trajectory quickly, which takes about 5 $\mathrm{s}$ in Fig.~\ref{fig8}. It also illustrates that the algorithm can solve the continuous-time finite-horizon control problem with the linear systems.

Lateral position control accuracy $I_{y{ \mathrm{err}}}$ is defined as the root-mean-square value of the error between the actual position and desired position; meanwhile, $I_{y\max }$ represents maximum lateral position error. They are formulated as:
\begin{equation}\begin{aligned}
I_{y{\mathrm{err}}}&=\sqrt{\frac{1}{N} \sum_{i=1}^{N}\left[y(i)-y_{\mathrm{des}}(i)\right]^{2},} \\
I_{y \max }&=\max _{i}\left\{\left|y(i)-y_{\operatorname{des}}(i)\right|\right.\}.
\end{aligned}\end{equation}

Lateral heading angle control accuracy and maximum heading angle error can be denoted as:
\begin{equation}\begin{aligned}
I_{\vartheta \operatorname{err}}&=\sqrt{\frac{1}{N} \sum_{i=1}^{N}\left[\vartheta(i)-\vartheta_{\operatorname{des}}(i)\right]^{2}}, \\
I_{\vartheta \max }&=\max _{i}\left\{\left|\vartheta(i)-\vartheta_{\operatorname{des}}(i)\right|\right.\}.
\end{aligned}\end{equation}

Vehicle comfort can be evaluated as:
\begin{equation}I_{y{\mathrm{comf}}}=\sqrt{\frac{1}{N} \sum_{i=1}^{N} r(i)^{2}}.\end{equation}

Here we evaluate the control performance using these indictors compared with MPC. The accuracy is similar, but the solving speed is greatly improved.
\begin{table}[htbp]
\caption{Comparison Of Control Performance Evaluation}
\begin{center}
\begin{tabular}{|p{1cm}|p{1cm}|p{1cm}|p{1cm}|p{1cm}|p{1cm}|}
\hline & $I_{y{\text {err }}}$ $(\mathrm{m})$ & $I_{y \max }$ $(\mathrm{m})$ & $I_{\vartheta \text { err }}$ $(\mathrm{rad})$ & $I_{\vartheta \max }$ $(\mathrm{rad})$ & $I_{\text {ycomf }}$ $(\mathrm{rad} / \mathrm{s})$ \\
\hline $\mathrm{ADP}$ & 0.013 & 0.52 & 0.0006 & 0.1338 & 0.025 \\
\hline $\mathrm{MPC}$ & 0.008 & 0.2 & 0.0002 & 0.043 & 0.008 \\
\hline
\end{tabular}
\end{center}
\end{table}

In this section, the simulation demonstrates that the proposed CT finite-horizon ADP algorithm can converge to the optimal policy for the linear system. Due to it is calculated off-line, it can improve efficiency.
\subsection{Discussion on problems of nonlinear dynamics}

\begin{figure}[htbp]
\centerline{\includegraphics[width=0.75\linewidth]{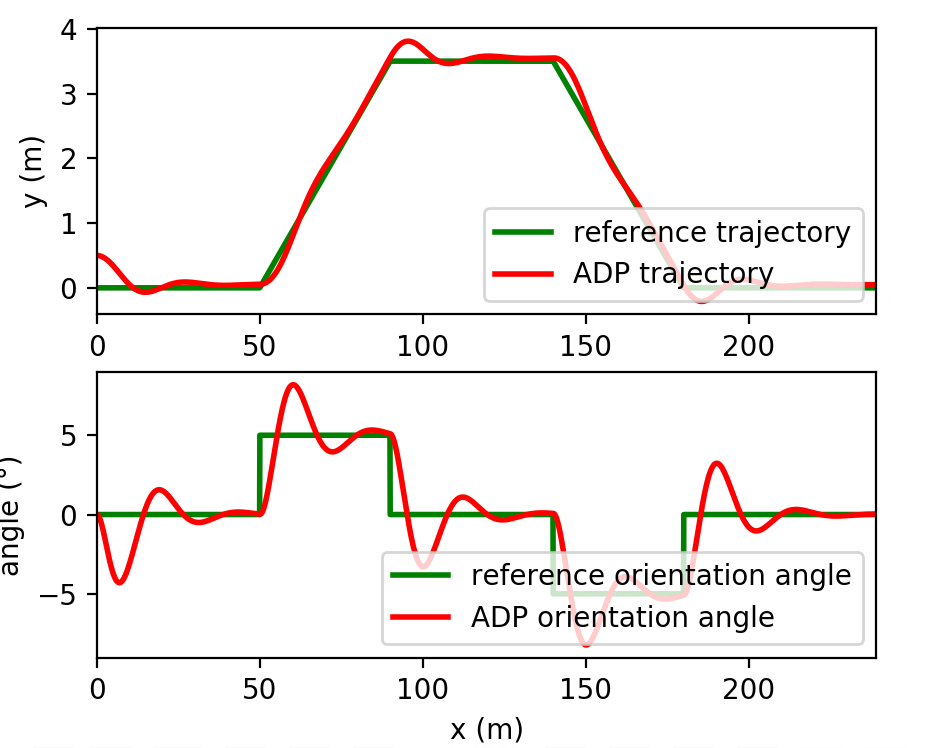}}
\caption{Tracking performance with nonlinear dynamic}
\label{fig9}
\end{figure}

We apply the CT finite-horizon ADP algorithm on a non-input-afﬁne nonlinear vehicle system derived as in \cite{kong2015kinematic}\cite{li2017driver}. The algorithm can also achieve excellent tracking performance, shown in Fig.~\ref{fig9}. It demonstrates that the proposed ADP algorithm can also be applied to nonlinear systems effectively.

\begin{figure}[htbp]
\centerline{\includegraphics[width=0.7\linewidth]{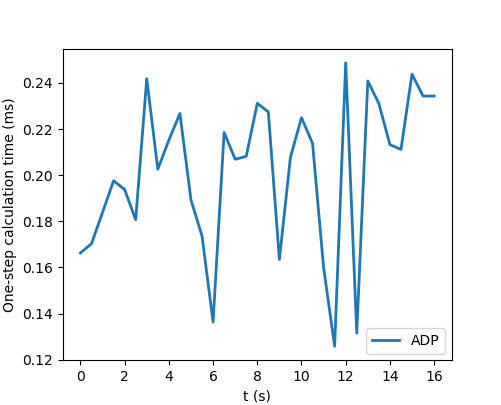}}
\caption{Nonlinear ADP calculation time}
\label{fig10}
\end{figure}

\begin{figure}[htbp]
\centerline{\includegraphics[width=0.7\linewidth]{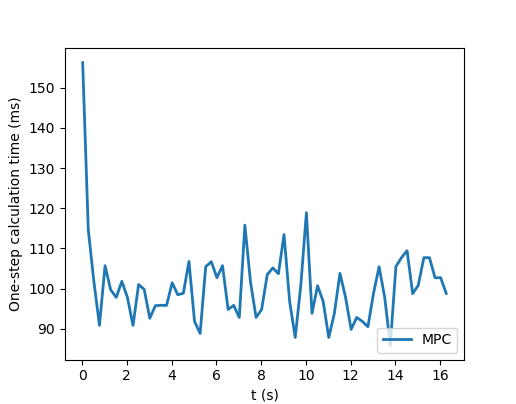}}
\caption{Nonlinear MPC calculation time}
\label{fig11}
\end{figure}

The simulation aims at control problems with finite-horizon $T=0.5$ $\mathrm{s}$. The frequency is 200 $\mathrm{Hz}$. It is obvious that QP solvers listed before are not suitable for this nonlinear problem. We compare the proposed ADP algorithm with ipopt solver in CasADi framework, which is a typical open-source software for nonlinear optimal control \cite{andersson2019casadi}. Our simulation is carried on personal computer with i7-8850H CPU. The results show that the average calculation time of ADP algorithm is 0.2 $\mathrm{ms}$ in Fig.~\ref{fig10}, yet nonlinear MPC requires nearly 101 $\mathrm{ms}$ in Fig.~\ref{fig11}.  It is easy to conclude that 
the proposed ADP algorithm is almost 500 times faster than nonlinear MPC algorithm.
\section{CONCLUSION}

The paper presents the continuous-time finite-horizon ADP algorithm for solving the general optimal control problem with known linear and nonlinear dynamics. We further verify the efficiency and optimality by the simulation on vehicle tracking problems. The result shows that it can calculate the off-line control policy for vehicle tracking problems with guaranteed optimality. For the linear control problem, the policy error is less than 1$\%$. For the nonlinear problem, the one-step calculation time of ADP is 500 times faster than the nonlinear MPC ipopt solver.
\section*{Acknowledgment}
We would like to acknowledge Zhengyu Liu, Hao Sun and Yarong Wang for their valuable suggestions for this research.


\bibliographystyle{IEEEtran}
\bibliography{icus2020}

\end{document}